\documentclass[preprint,proceedings]{rmaa} 
 
\usepackage{paralist} 
 
\usepackage{psfrag,color} 
 
\newcommand{\cms}{\hbox{${\rm cm^{-2}}$}}

\newcommand{\anu}{\hbox{$\alpha_{\nu}$}}

\newcommand{\zq}{\hbox{$z_q$}}

\newcommand{\HI}{\hbox{H\,{\sc i}}}

\newcommand{\Lya}{\hbox{Ly$\alpha$}}

\newcommand{\Nh}{\hbox{$N_{H}$}}

\SetYear{2005} 
\SetConfTitle{Ninth Texas-Mexico Conference on Astrophysics} 
 
\title{Dust and the ultraviolet energy distribution of quasars}  
 
\author{ 
  Luc Binette,\altaffilmark{1}  
  Christophe Morisset\altaffilmark{1}
  and Sinhue Haro-Corzo\altaffilmark{1}} 
 
\altaffiltext{1}{Instituto de Astronom\'\i{}a, UNAM, M\'exico D.F.,  
M\'exico.} 
 
\shortauthor{Binette et al.} 
\shorttitle{Dust and the spectral energy distribution of quasars} 
 
\fulladdresses{ 
\item  Luc Binette, Christophe Morisset and Sinhue Haro-Corzo: Instituto de  
Astronom\'\i{}a, Universidad Nacional 
  Aut\'onoma de M\'exico, Ciudad Universitaria, Apartado Postal 70--264, CP 
  04510, M\'exico D.F., M\'exico (\email{binette@astroscu.unam.mx}). 
} 
 
\listofauthors{L. Binette, C. Morisset, S. Haro-Corzo} 
\indexauthor{ Binette, L.} 
\indexauthor{Morisset, C.} 
\indexauthor{Haro-Corzo, S.} 
 
\abstract{The ultraviolet energy distribution of quasars shows a sharp
steepening of the continuum shortward of  1000\,\AA\ (rest-frame). We
describe how we came to consider the possibility that this 
continuum break might be the result of  absorption by carbon crystallite
dust grains.}

\resumen{La distribuci\'on de energ\'{\i}a de los cuas\'ares presenta en el ultravioleta
un quiebre abrupto a 1000\,\AA.  Se describe como
llegamos a considerar que el quiebre se deb\'{\i}a a la absorci\'on
por granos de polvo constituidos de carbono cristalino.}
 
\addkeyword{galaxies: intergalactic medium } 
\addkeyword{galaxies: active} 
\addkeyword{radiative transfer} 
\addkeyword{ultraviolet: general}

\begin{document} 

\maketitle 
 
\begin{figure*}[!t]
  \includegraphics[width=\columnwidth,height=8cm]{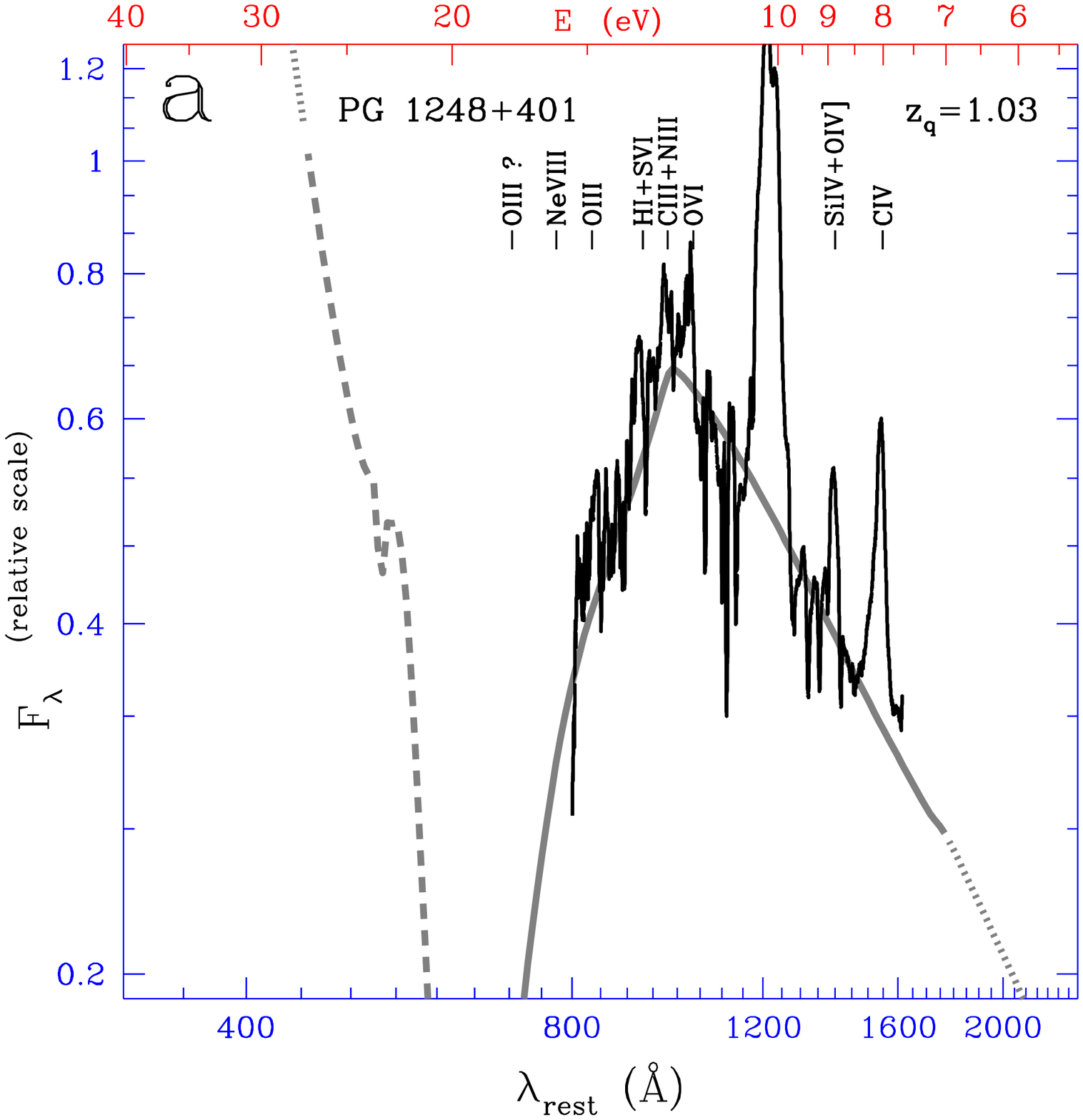}%
  \hspace*{\columnsep}%
  \includegraphics[width=\columnwidth,height=8cm]{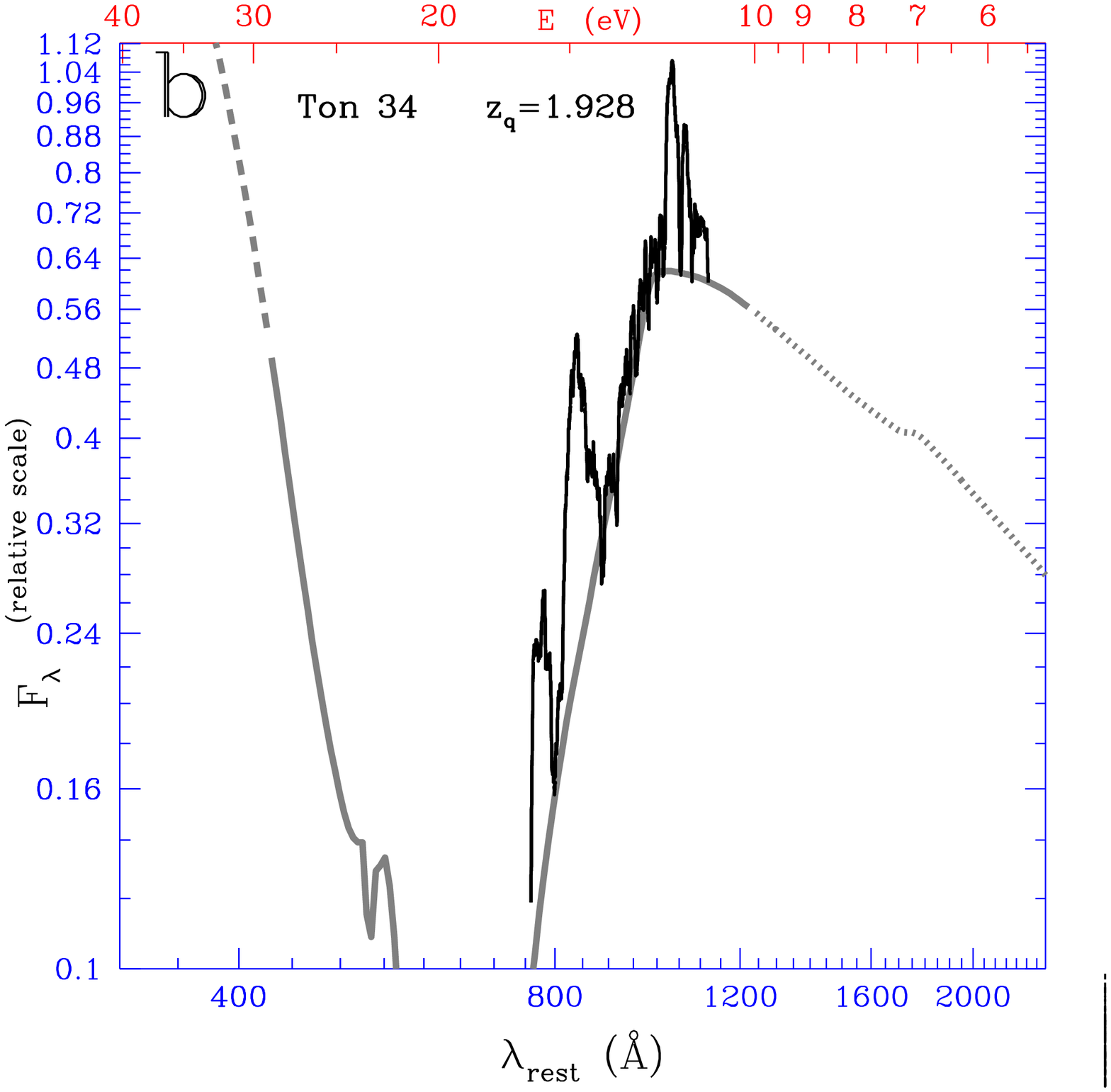}
  \caption{Panel a: rest-frame spectrum of PG\,1248+401 in black. The gray line
  represents a dust absorption model with $\Nh =3.2\times
  10^{20}$\,\cms\ using cubic diamond grains.  Panel b: rest-frame spectrum of
  Ton\,34 in black. The gray line represents a dust absorption model
  with $\Nh=5.0 \times 10^{20}$\,\cms\ using cubic diamond grains.}
  \label{fig:classb}
\end{figure*}

\section{Introduction} \label{sec:intro} 
 
The ultraviolet energy distribution of quasars is characterized by the
so-called ``big blue bump'', which peaks in ${\nu}F_{\nu}$ at
approximately 1000\,\AA.  The quasar `composite' spectral energy
distribution (SED) of Telfer et\,al. (2002, hereafter TZ02) obtained
by co-adding 332 HST-FOS archived spectra of 184 quasars between
redshifts 0.33 and 3.6, exhibits a steepening of the continuum at
$\sim 1100$\,\AA. A fit of this composite SED using a broken powerlaw
reveals that the powerlaw index changes from approximately $-0.69$ in
the near-UV to $-1.76$ in the far-UV. We label this observed sharp
steepening the `far-UV break'. In these proceedings, we describe how we came to 
propose that absorption by crystalline carbon  dust is the possible cause 
of the UV break observed in high redshift  quasars. 
The argumentation behind this interpretation of the UV break 
has been presented in detail in Binette et\,al. (2005a, hereafter BM05). 
Further information  can be found in recent proceedings such as in Binette et\,al. (2005b, c).

\section{The nearby AGN with FUSE} \label{fuse}

In an earlier paper, Binette et al. (2003) showed that \HI\ scattering
by a tenuous intergalactic component could {\em not} be the cause of
the 1000\,\AA\ break. This negative result supported the prevailing
view that the break is an intrinsic feature of quasar SED.  More
recently, Scott et\,al. (2004) derived a composite SED similar to TZ02
but for `nearby' ($\zq<0.7$) active galactic nuclei (AGN), using
archived data from FUSE. The authors reported the lack of evidence of
a steepening in nearby AGN! This new piece of information intrigued
us.  Although this absence of a continuum break could be explained
away by supposing that the nearby AGN are less luminous and hence
possess a lower mass blackhole and as a result a hotter accretion
disk, we were not initially satisfied by this explanation. An
additional reason for being skeptical is that some individual spectra
from the TZ02 sample are extremely far-UV deficient, showing a much
steeper break than that seen in the composite. Two examples are given
in Fig.\,\ref{fig:classb}.  Yet their emission-line spectrum is no
different than that of other quasars. Photoionization is generally
believed to be the excitation mechanism of the emission lines.
Therefore, the above-mentioned UV deficiency poses a serious challenge to
our understanding of what mechanism powers the emission lines.

The original suggestion of investigating dust absorption as a possible
cause of the break came from one of us, C. Morisset (CM), who had
experimented with photoionization models of planetary nebulae that
included dust mixed with the ionized gas.  CM's suggestion arose after
looking at an interesting figure prepared by S. Haro-Corzo (SHC), in which
three spectra appeared, showing a steep far-UV break. LB argued that ISM
dust could not reproduce the sharpness of the 1000\,\AA\ break, as had already
been  shown by Shang et al. (2004). This initial suggestion
nevertheless stayed around on and lead to a bibliographical search by LB of a
new  grain composition, which would have  the property of producing a sharp break at
1000\,\AA.

\section{Comparing radio-loud and radio quiet quasars}\label{sec:rqrl}

The first step has been to explore whether there might be evidence of
reddening within the quasar sample that Telfer kindly lent to us
in 2002. If dust was responsible for the break, we might for instance
expect that the degree of steepening would scale with the amount of
dust present. TZ02 had previously showed that the far-UV continuum was steeper 
in radio-loud (RL) than in radio-quiet (RQ)
quasars.  Within the paradigm of the dust being the cause of the
break, this difference must be the result of differences in the
amount of dust present. In other words, radio-loud quasars are
possibly more absorbed\footnote{This difference in absorption
might be statistical in nature and arise from the particular 
RL and RQ subsamples at hand.} than radio-quiet quasars.  

\begin{figure}[!t]
  \includegraphics[width=\columnwidth]{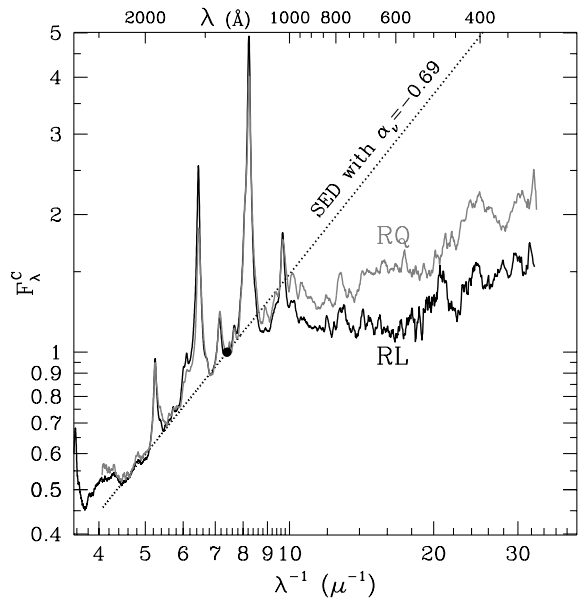}
  \caption{The composite spectral energy distributions of Telfer et\,al. (2002) for 
 radio-quiet (gray: RQ) and radio-loud (black: RL) quasars, respectively.}
  \label{fig:sed}
\end{figure}

To verify this proposition, we over-plot in Fig.\,\ref{fig:sed}
 the separate radio-quiet and  radio-loud composite SEDs derived by TZ02.
Each composite in this figure, however, has  been multiplied by the
appropriate normalization constant that made their flux equal to unity at
1350\,\AA. The radio-loud and radio-quiet
composites in Fig.\,\ref{fig:sed} are painted in black and gray, respectively. The black dot
represents the renormalization wavelength.  Within the narrow spectral
segments that appear to be line-free between 2500 and 1200\,\AA, both
continua   overlap remarkably well. This suggests that the
intrinsic SED longward of \Lya\ are very similar in  both quasar subsets. The dotted line in
panel {\it a} is a powerlaw fit to  line-free segments, using the
 mean spectral index value of $\anu = -0.69$ determined by TZ02
for the combined RL+QR sample.

\begin{figure}[!t]
  \includegraphics[width=\columnwidth]{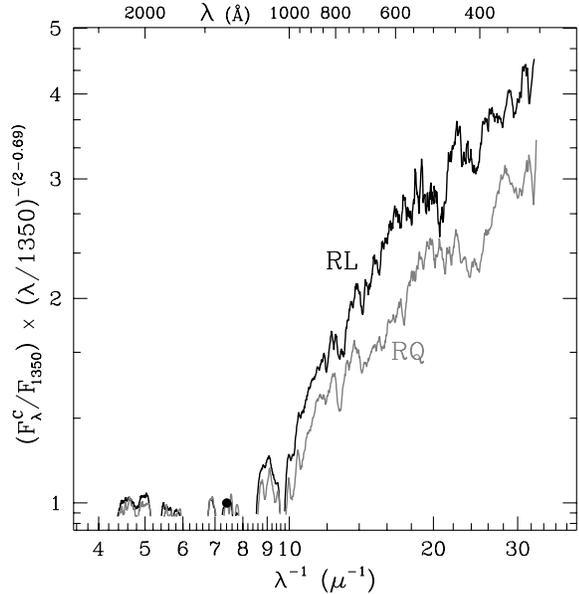}
  \caption{The ratio of the fitted powerlaw with each   composite SED:  
 radio-quiet  quasars (gray)  and radio-loud quasars (black).}
  \label{fig:abs}
\end{figure}

If we make the simplification that dust absorption is negligible below
1000\,\AA, we can take the ratio of the fitted powerlaw with either
composite SED as a means of showing how the UV deficit increases with
wavelength.  Such ratio-curves are plotted in Fig.\,\ref{fig:abs}. We
can see that the UV deficit increases smoothly, starting at the break,
near 1000\,\AA, down to about 550\AA. The UV deficit increases faster
in the case of radio-loud quasars than in radio-quiet quasars.  Such a
difference in the behavior of the ratio-curves is expected, if RL
quasars are more absorbed than RQ quasars, and dust absorption is
responsible for the UV deficit. The absorption would be characterized
by an absorption cross-section that increases toward shorter
wavelengths. At wavelengths shorter than 550\,\AA, some spectral
features appear to be unique to each composite. The absorption test
becomes therefore inconclusive in that region. This could be the
result of having too few very high redshift quasars among the TZ02
sample, which  result in a loss of reliability of the composite
SEDs in that wavelength domain (see TZ02).

\section{Carbon crystallites}  \label{sec:C}

The UV deficit in RL and RQ quasars has been shown to behave qualitatively 
as expected,
if dust absorption was responsible for the  break. 
The next step  consisted in searching the kind of 
material that might possess the optical properties   
required to produce a sharp absorption 
feature at the same position as the 1000\,\AA\ break. 
We looked for a dust constituent, whose absorption cross section 
peaked in the far-UV and yet causes negligible absorption at wavelengths
longer than 1200\,\AA. Ideally, as it is the case with the interstellar medium (ISM) dust, 
the grain
particles should be composed of the most abundant elements. Using ADS and Google, 
the most promising candidate appeared to be carbon in its crystal form, 
but with surface impurities. The so-called meteoritic nanodiamonds.
The finding of the  the recently published work on nanodiamonds by
Mutschke, Andersen 
and coworkers\footnote{Both Harald Mutschke and Anja Andersen have since 
become collaborators of this project.} (Mutschke et\,al. 2004) 
lead to a real breakthrough in the project, 
since  Mutschke et\,al. (2004) had just measured the optical properties 
of nanodiamonds down to very short wavelengths. The grain size distribution 
could in principle be varied as needed, using the Mie theory to compute the 
extinction curve  (Binette et\,al. 2005a, b). 
It turns out that it is unnecessary 
to assume a different 
grain size distribution than that which is found to characterize nanodiamonds
embedded in  primitive meteorites [Lewis et\,al. 1989] 
(provided the dust is  intrinsic to quasars and not extragalactic, 
see BM05).

\section{Dust grains with and without surface impurities} \label{sec:imp}

As matters stand, the crystalline form of carbon can exist 
either in the form of the well known terrestrial type of cubic diamonds  
or as the type found in primitive meteorites as 
in the Allende\footnote{This meteorite belongs to the category of 
carbonaceous chondrite meteorites. It is a
relatively rare meteorite type, at a frequency of only $\sim 3.5$\%. The particular 
Allende
meteorite who fell on Earth near the town of Allende in the state of
Chihuahua, M\'exico, on February 8th, 1969, is one of the most studied
meteorites of its kind.}  meteorite, which was incidentally 
used by Mutschke et\,al. (2004) 
in their study of non-terrestrial nanodiamonds.

\section{Results}\label{sec:res}

\begin{figure}[!t]
\includegraphics[width=\columnwidth]{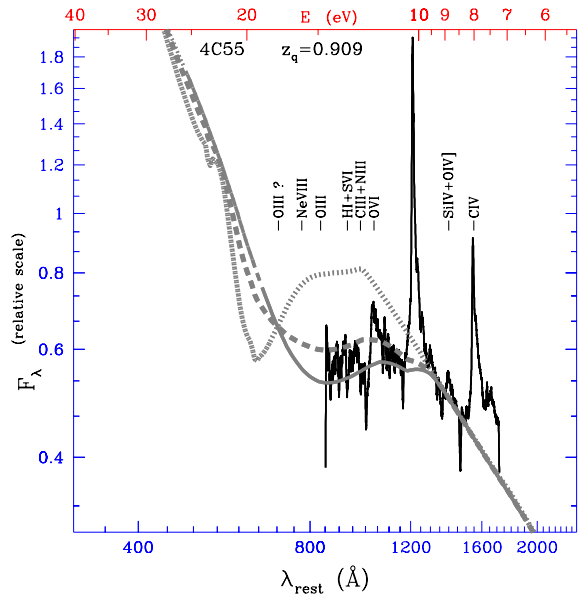} \caption{The
  spectral energy distributions of 4C55. Gray continuous line:
  absorption model with $\Nh=1.1 \times 10^{20}\,$\cms\ using
  nanodiamond grains from the Allende meteorite.  Dashed line: absorption model
  (same column) using a grain mixture of 30\% cubic diamonds and 70\%
  Allende nanodiamonds. Dotted line: absorption model (same column)
  using cubic diamonds only.  } \label{fig:classa}
\end{figure}

Using the complex refraction indices $n+ik$ from Mutschke et\,al
(2004) for the Allende nanodiamonds, and Edwards \& Philipp (1985) for
the cubic diamonds, respectively, we proceeded to calculate the
extinction curve corresponding to each of the two nanodiamond
types. Assuming a simple intrinsic SED consisting of a powerlaw with
the spectral index inferred from the observed near-UV region in each
quasar, we proceeded to calculate the absorbed powerlaw and compare it
with the observed SED. We found that an acceptable fit could be
obtained of the 1000\,\AA\ break in 80\% of quasars. However, the
dust absorbed powerlaw model requires in most cases  that dust
grains of the above two types be combined  (the terrestrial cubic diamonds and the
nanodiamonds of the type found in primitive meteorites). This result, as well as the
computed extinction curves are shown  in the proceedings of
another meeting (Binette et al. 2005b). We will present here only a
few  spectra that  can be fitted, using a {\it single} type of nanodiamonds.

As shown in
Fig.~\ref{fig:classb}, the cubic diamond extinction curve fits the abrupt breaks
found in the quasars PG\,1248+401 and Ton\,34 very well. 
The Allende nanodiamonds, on the other hand fit
better the break observed in 4C55 (continuous line), as shown in
Fig.\,\ref{fig:classa}, where a comparison is also made with pure
cubic diamonds (dotted line) or a combination of the two nanodiamond types
(dashed line). The hydrogen columns quoted in the figure captions
assume that all  carbon is locked unto the dust, and that its
abundance is solar.  It corresponds to a dust-to-mass ratio of
0.003. The real dust-to-gas ratio cannot be constrained at this stage.

Instead of using
optically known materials, one could have treated the
absorption hypothesis as an inverse problem, working out the
extinction curve that best succeeds. We consider, however,  that it confers a  higher
degree of plausibility to have used an extinction curve based on a known material
such as that of the
Allende meteorite, rather than an invented cross-section. Finally,
the vector that we propose to be responsible for the
absorption consists of grains made of carbon atoms, a major
constituent of the interstellar medium dust, albeit here in a less common  form, that of
crystals (nanodiamonds).

\acknowledgements 
 
The authors acknowledge support from CONACyT grant 40096-F. 
We thank Randal Telfer for sharing the reduced HST FOS spectra.
Diethild Starkmeth helped us with proof reading.

\end{document}